# White light generation and anisotropic damage in gold films near percolation threshold

Sergey M. Novikov[1], Christian Frydendahl[2,3], Jonas Beermann[1], Vladimir A. Zenin,[1], Nicolas Stenger[2,3], Victor Coello[4], N. Asger Mortensen[2,3] & Sergey I. Bozhevolnyi[1]

[1]*Centre for Nano Optics, University of Southern Denmark, Campusvej 55, DK-5230 Odense M, Denmark*
[2]*Department of Photonics Engineering, Technical University of Denmark, DK-2800 Kongens Lyngby, Denmark*
[3]*Center for Nanostructured Graphene, Technical University of Denmark, DK-2800 Kongens Lyngby, Denmark*
[4]*CICESE Monterrey, Alianza Centro No.504 PIIT Apodaca, N. L. C. P. 66600, Mexico*

**Strongly enhanced and confined electromagnetic fields generated in metal nanostructures upon illumination are exploited in many emerging technologies by either fabricating sophisticated nanostructures or synthesizing colloid nanoparticles. Here we study effects driven by field enhancement in vanishingly small gaps between gold islands in thin films near the electrically determined percolation threshold. Optical explorations using two-photon luminescence (TPL) and near-field microscopies reveals super-cubic TPL power dependencies with white-light spectra, establishing unequivocally that the strongest TPL signals are generated with close to the percolation threshold films, and occurrence of extremely confined (~ 30 nm) and strongly enhanced (~ 100 times) fields at the illumination wavelength. For linearly polarized and sufficiently powerful light, we observe pronounced optical damage with TPL images being sensitive to both wavelength and polarization of illuminating light. We relate these effects to thermally induced morphological changes observed with scanning electron microscopy images. Fascinating physics involved in light interaction with near-percolation metal films along with their straightforward and scalable one-step fabrication procedure promises a wide range of fascinating developments and technological applications within diverse areas of modern nanotechnology, from bio-molecule optical sensing to ultra-dense optical data storage.**

**Introduction**

Illumination of metal nanostructures results in nanostructured optical fields that are strongly enhanced and localized in the vicinity of sharp corners and in nanometer sized gaps between metal surfaces[1]. While diverse field enhancement (FE) effects can be realized by dedicated design and high-resolution (often rather sophisticated) nanofabrication[2], intriguing optical properties including FE-driven linear and nonlinear effects can be found in thin semi-continuous films near the percolation threshold[3] – the critical point at which individual metal clusters start forming connected structures across the substrate domains[4–7]. These films can be obtained by simple and straightforward metal deposition, e.g. thermal evaporation, onto a dielectric or semiconductor substrate. As the average film thickness increases during the metal deposition, individual and well separated metal islands extend their sizes, eventually forming near the percolation threshold a semi-continuous film that gradually evolves into a homogeneous smooth film exhibiting (close to) bulk properties[7].

The percolation threshold can experimentally be observed by monitoring FE effects with optical methods[3–5] and uniquely determined in the low-frequency regime through electric conductivity measurements[8,9]. It should also be noted that the average film thickness corresponding to the percolation threshold depends strongly on the deposition conditions, substrate material and metal involved[10, 11]. Considerable interest in semi-continuous metal films near the percolation threshold is primarily motivated by their remarkable ability of generating (under illumination) strongly enhanced local electromagnetic fields, so-called 'hot-spots'[12–15], that can be observed directly with near-field microscopy techniques[5] and indirectly via strongly enhanced nonlinear optical interactions[3,7,8]. Typically, hot spots are associated with FEs occurring in nm-sized gaps between metal surfaces of nanoparticles (NPs), clusters or specifically designed nanostructures, with nonlocal effects becoming significant for sub-nanometer gaps where they limit the FE levels[16, 17] with far-reaching implications also for the nonlinear dynamics[18–21].

Strong FE effects in metal nanostructures primarily occur due to resonantly excited surface plasmons (SPs), i.e. collective electron oscillations in metals coupled to electromagnetic fields in dielectrics[1-3,22–24]. Resonant interactions in metal nanostructures involving localized as well as propagating SPs have been investigated using colloidal metallic NPs of various sizes and shapes[24–27], NP ensembles[28,29] with pre-determined optical properties, periodic[30–32] and random[33] nanostructures. The spectral position of resonances is tunable through a variety of parameters such as geometry, composition of nanostructures or size and shape of NPs[23-25,34]. These artificial nanostructures and NPs represent well-defined regular configurations exhibiting resonant FE at one or several wavelengths[25–28,30,31] or irregular random nanostructures hosting (spatially separated) resonant excitations covering a wide spectral range[33]. One signature of extremely strong electric fields in metal nanostructures is white-light generation, first observed with resonant optical nano-antennas[35].

Strong FEs are extremely important for both fundamental studies within an emerging field of quantum plasmonics[4,36–38] and practical applications such as sensors[25,27], playing a major role in surface-enhanced spectroscopies, including surface-enhanced Raman scattering (SERS)[30,39–41]. Another interesting application of strong FE effects is pulsed-laser induced colour changes of aluminium disk-hole resonances[42] and optical recording in gold nanorod solutions mediated by local plasmonic heating and melting[43,44]. It should however be noted that widely used colloids of NPs, e.g., nanorods, could aggregate during their deposition on a sample surface, leading to strong variations in optical properties over the sample surface and thereby decreased reproducibility. At the same time, electron-beam lithography (EBL) and focused ion-beam (FIB) milling techniques, that do offer high reproducibility, feature limited (~10 nm) spatial resolutions, limiting thereby achievable FE levels, and require rather costly equipment, thus hindering low-cost and large-scale production. Alternatively, semi-continuous metal films near the percolation threshold can quite easily be fabricated by evaporation of metals like gold or silver onto a dielectric or semiconductor substrate. Using electron-beam deposition it is possible to fabricate structures covering rather large, up to wafer size, areas in only a few minutes.

Two-photon luminescence[45,46] (TPL) microscopy is a powerful tool for characterizing FE effects[47], and we have recently observed significant FE effects by conducting TPL and SERS experiments with thin gold films during their transition from low-coverage (island-like) to continuous films[7]. In the present work, by carefully mapping this transition with parallel TPL microscopy and electrical conductivity characterization, we demonstrate unequivocally that TPL signals reach maximum levels at the percolation threshold, featuring super-cubic power dependencies with white-light spectra similar to those obtained with extreme FEs in plasmonic nanoantennas[35]. Moreover, for linearly polarized and sufficiently powerful light, we observe pronounced optical damage with subsequent TPL images being sensitive to both wavelength and polarization of the illuminating light. We relate the polarization and wavelength sensitivity of optical damage to thermally induced morphological changes (at locations of extreme FEs) observed with scanning electron microscopy (SEM). Finally, using phase- and amplitude-resolved near-field imaging of near-percolation films, we directly demonstrate the occurrence of extremely confined and enhanced fields at the illumination wavelength with the FE being both polarization and wavelength dependent. Near-field images allow us to elucidate the underlying physical mechanisms involved in the observed TPL phenomena and evaluate the high FE levels achieved in our experiments.

**Results**

**TPL enhancement and white light generation.** Thin gold films with thicknesses ranging from 2 to 11 nm were deposited onto room-temperature glass substrates (see Methods). A subset of films (with thicknesses of 3, 5, 7, and 9 nm) was also prepared on 18 nm-thin $SiO_2$ membranes for transmission electron microscopy (TEM) imaging (Fig. 1a-d). It should be noted that the thickness of gold films specified throughout this work is the average nominal coverage measured by a quartz oscillator, and variations of the order of ±0.5 nm across a 4" wafer are expected. Very thin films with thicknesses of up to 3 nm consist of well-separated islands (Fig.1a), whereas those with larger thicknesses, from 5 to 7 nm, feature labyrinthine structures (Fig. 1b,c). Finally, films become practically continuous for the thickness of 9 nm (Fig. 1d). Prior to TPL investigations, the fabricated samples are characterized by linear reflection spectroscopy. As expected intuitively, the film reflectivity increases gradually and monotonously with nominal film thickness, without revealing any specific resonances in the frequency range of interest (Supplementary Fig. 1a). Characterization of the FE effects in thin gold films is then conducted using the TPL microscopy and spectroscopy (see Methods). In passing we note that, in all configurations considered here, TPL signals disappear completely after switching a pump laser from fs-pulsed to continuous operation mode, a clear signature of the non-linear origin of the investigated phenomenon.

To compare thin films with bulk gold, we fabricated a complimentary set of samples, where thin gold films are evaporated onto a glass substrate, half of which was initially coated (by evaporation) with a 100-nm-thin gold film, referred hereafter to as bulk gold. For simplicity, we kept the same excitation wavelength of 740 nm in these measurements, since there are no specific resonances at this wavelength and 740 nm is better visible and convenient for alignment. Cross sections of scanning optical microscopy maps across the boundary between bulk gold and thin gold films, obtained simultaneously with a band-pass filter at the illumination wavelength (noted as the fundamental harmonic, FH) and with a low-pass filter transmitting only the TPL signals, reveal significantly different optical responses (Fig. 1e). As expected, the level of FH signals from the bulk gold is the same for all samples, whereas the level of FH signals from thin films decreases with a reduction of thickness (in accordance with the thickness-dependent linear optical spectra shown in Supplementary Fig. 1a). Contrary to that, the TPL signals from thin films are significantly higher than those from the bulk gold, depending non-monotonically on the film thickness (in accordance with the thickness-dependent TPL spectra shown in Supplementary Fig. 1c,d). The strongest TPL signals are observed with the 5-nm-thin film, being ~10 times higher than those for the 9-nm-thin film, which exhibits a very weak and rather homogeneous TPL response (as expected for thick films approaching bulk gold in their optical properties). The TPL response from the 5-nm-thin film features noticeable oscillations at the level of ~ 10% (Fig. 1e.), indicating that the TPL sources, which are expected to originate from subwavelength-sized hot spots, differ in strength and are randomly distributed over the film area. These features are also corroborated with near-field optical images and their analysis (Supplementary Discussion 1).

As the next step, we compare the average TPL signals measured with thin films of different thicknesses with the resistance measurements (see Methods) of the same films (Fig. 2a). One can conjecture that the percolation threshold in electrical conductivity occurs for the film thickness being between 4 and 5 nm, since for a 4-nm-thin and thinner films we could not observe any finite electrical conductance. The absence of percolation for very thin films is also supported with the TEM images (Fig. 1a,b). By comparing the electrical measurements with the TPL data obtained at the same illumination conditions for all film thicknesses, we conclude unequivocally that TPL signals reach maximum levels at the percolation threshold (Fig. 2a). Such correlation is reasonable, since the largest density of the small gaps, at which hot spots could form, is expected near the percolation threshold. With further gold deposition small gaps close, causing gold islands to merge into pathways and eventually approach a continuous film (Fig. 1d).

The TPL signals, originating from the two-photon absorption, are expected to be proportional to the square of the incident power[45-47]. It turned out, however, that, for all thin films, the TPL signals feature super-cubic power dependences (Supplementary Fig. 1b). A possible explanation of this is supercontinuum white-light generation in the sample[35,48], possibly driven by non-local electron response[49] in complex geometric shapes of semi-continuous plasmonic nanostructures[21]. The white-light generation indicates the occurrence of strong FE effects, responsible for giant local fields, at not only illumination but also photoluminescence wavelengths. The broadband nature of FE effects is in fact expected for semi-continuous films near the percolation threshold[3,6,33]. To confirm the white-light continuum generation we recorded luminescence spectra (see Methods) for the same film thicknesses as those used in the measurements shown in Fig. 1e, but at the excitation wavelength of 780 nm (Fig. 2b). For comparison, we also recorded luminescence spectra for another excitation wavelength, 740 nm, and other film thicknesses (Supplementary Fig. 1c,d). All spectra measured with thin films appear similar (apart from the signal level) and very different from the spectrum recorded with the 100-nm-thick gold film, especially in the wavelength range of 600-700 nm. The thin-film spectra feature two broad maxima near 550 and 675 nm, which could be interpreted in terms of discrete transitions between electronic states in gold and influenced by localized SP resonances. It should be mentioned, that the photoluminescence was so strong that it was even possible to directly observe it by eye in the microscope when scans were not recorded. Note that white-light continuum generation in gold was first observed also in the range near 550–600 nm[35].

**Photo-thermally induced anisotropic damage.** It is intuitively expected that the TPL cannot steadily increase with an increase of the incident illumination power simply because any optical absorption is accompanied by heating. In our case, this heating causes eventual damage of the metal nanostructures by their melting and reshaping. The photo-thermally induced damage incurred by scanning a sample area with a focused pump beam, hereafter referred to as a "writing" process, can easily be visualized in a "reading" procedure, when a larger surface area is subsequently scanned with a laser power well below the damage threshold. The same procedure can also be used to verify the absence of damage (and it was systematically used in our experiments reported here). The TPL from the written pattern appears to be weaker than that from the surrounding undamaged area, with TPL signals from written areas decreasing rapidly when increasing the laser power used for writing (Supplementary Fig. 2). This effect can be explained by local heating, melting and reshaping of gold nanostructures at hot spots, since strong enhancement of local fields implies strong (and local) enhancement of absorption of radiation due to Ohmic losses. It should be noted that the damage could not be reproduced with the laser operating in the continuous mode with the same average power, an important observation indicating that the heating primarily occurs in the process of two-photon absorption, with the energy difference (between two photons absorbed and one emitted) being spent to heating. Additionally, it is expected that the temperature rise is significantly weaker in the continuous operation mode due to fast heat dissipation from relatively small heated volumes of hot spots. Interestingly, a decrease in TPL signals from the damaged area was only observed when the laser polarization was the same during both reading and writing, while for orthogonal polarizations TPL signals from the damaged area were practically the same or even higher than those from the neighboring undamaged areas (Supplementary

Fig. 2). However, such anisotropy was not well reproducible and clearly pronounced for thin gold films on glass substrates. It is reasonable to suggest that thermal conductivity of a substrate plays an important role in the optical damage by influencing the rate of heat dissipation from strongly localized hot spots to the substrate.

In order to elucidate the influence of the substrate material we prepared gold films with nominal thicknesses of 3, 5, 7, and 9 nm on high resistivity silicon substrates (see Methods). It turned out, for the same incident power, that the TPL from thin gold films on silicon substrates is considerably weaker, practically by two orders of magnitude, than the TPL from similar thin films on glass substrates, although the luminescence spectra are very similar (Supplementary Fig. 3). We relate this striking difference to a very large difference in the thermal conductivity of glass and silicon, which is also about two orders of magnitude[50], because relaxation processes in gold are expected to speed up at elevated temperatures. At the same time, the polarization anisotropy effect in writing and reading out experiments, although requiring large powers of illumination, became much more pronounced and better defined compared to that observed with the films on glass substrates. Two orthogonal pairs of stripes written with orthogonal polarizations can be observed separately and practically without cross talk by using the corresponding orthogonal polarizations during the read-out procedure (Fig. 3a-c). Noting that thermal conductivities of silicon and gold are of the same order of magnitude[50], we explain the observed differences by the circumstance that gold nanostructures on glass substrates are heated to higher temperatures and more homogeneously than those on silicon substrates, because the latter serve as a very efficient heat dissipation channel (even more efficient than thin gold films). Consequently, the damage in gold nanostructured films on silicon substrates is expected to be stronger localized, essentially to the area of hot spots.

The hypothesis is therefore that, within the illuminating 750-nm-diameter beam spot (see Methods), there are several bright (dipolar) localized SP excitations (hot spots), and each can be excited with a particular linear polarization, contributing substantially to the TPL signal obtained in the read-out procedure. It is also reasonable to expect that many hot spots are related to the gap-induced FEs with the electric fields being strongly enhanced in tiny gaps (oriented perpendicular to the incident field polarization) between resonant nanoparticle[35,51-53]. The hot spots can be damaged by local heating (when using silicon substrates) with subsequent reshaping but only when using the correspondingly polarized illumination during writing. Damaged locations can no longer efficiently contribute to TPL when the same polarization is used during reading, resulting in the overall decrease of the TPL signal. At the same time, for the read-out with the orthogonal polarization this damage can hardly be seen, since the damaged locations were not hosting hot spots for this polarization in the first place. This explanation accounts also for the polarization dependent damage observed in the FH images (i.e., at the illumination wavelength), although with a reduced resolution and contrast (Supplementary Fig. 4) simply because of linear contrast being weaker than the nonlinear one.

The suggested mechanism is somewhat resembling that exploited in the polarization and wavelength multiplexed optical recording mediated by SP in gold nanorods[43]. In our case, the TPL images obtained during the read-out procedure also exhibit the sensitivity with respect to the illumination wavelength: the contrast becomes stronger for longer wavelengths and (unexpectedly) inverted for short wavelengths (Fig. 3d-f). Moreover, the contrast inversion for short wavelengths is also observed in the FH images (Supplementary Fig. 5). These effects can be explained within the same hypothesis described above by taking into account the fact that metal nanoparticles upon melting tend to decrease their surface area due to surface tension[1]. Therefore, elongated nanoparticles that are resonant at a given wavelength become thicker and shorter during writing at this wavelength with their resonance shifting to shorter wavelengths. Their contribution to both FH and TPL signals are thereby increasing for shorter and decreasing for longer read-out wavelengths (Supplementary Fig. 5). The damage mechanism described above is consistent with SEM images of pristine and damaged areas of the 5-nm-thin gold film on a silicon substrate, showing unambiguously an increase of inter-particle distances along the polarization direction of writing field due to reshaping, including merging of tiny particles, caused by (enhanced and localized) heating induced by the gap-induced FEs (Fig. 4a,b). This observation accounts for the pronounced polarization effects in the TPL writing/reading procedures. Other damage effects include spatial subwavelength-sized localization of damaged areas and reshaping in the form of rounding of particles (Supplementary Fig. 6). The latter feature is important for understanding of the contrast inversion in the FH and TPL images at wavelengths shorter than that used for writing.

**Near-field imaging.** The proposed mechanism of the photo-thermally induced anisotropic damage relies on the existence of strongly enhanced and confined (dipolar) resonant SP excitations in near-percolation thin gold films used in our experiments. In principle, their existence at thin (semi-continuous) near-percolation metal films illuminated with practically any wavelength is well documented[3-6,12-15,33], but their spatial extensions and the corresponding FEs as well as the polarization and wavelength sensitivity are strongly dependent on the actual film morphology, e.g., self-similarity and self-affinity properties[33]. In order to reveal main features in spatial distribution of hot spots and associated FEs, we conducted near-field phase- and amplitude resolved mapping of local optical fields formed at the pristine 5-nm-thin gold film on a glass substrate illuminated with a tuneable continuous laser at telecom wavelengths (see Methods). Scanning near-field optical microscopy (SNOM) images revealed the existence of randomly distributed and strongly localized (~ 30-nm-wide) and enhanced (Supplementary Fig. 7) electromagnetic excitations with the FE levels that exhibit a well pronounced polarization dependence, indicating a dipolar response (Fig. 4c-f). For example, intense hot spots marked as "1" and "2" in Fig. 4d-e are lighting up for orthogonal polarizations of the incident light while being separated by ~ 400 nm, a separation which is smaller than the illuminating beam spot size in the TPL experiments. These observations strongly support our explanation of the polarization dependent TPL (and FH) writing and reading. Although the direct overlap of the simultaneously recorded hot spots with topography suggests that the hot spots observed can primarily be related to gaps between particles or to single particles (Supplementary Fig. 8), we find this evidence inconclusive because of a limited SNOM resolution (~ 10 nm) dictated by the probe tip size[54]. Near-field images obtained at different illumination wavelengths indicate that the FE levels at hot spots weakly depend on the wavelength, at least within the wavelength interval (~ 200 nm) available for near-field characterization (Supplementary Fig. 9). Note that the gap-induced FEs are associated with the boundary conditions for the electric field[51,52], and are thereby weakly wavelength dependent. Overall, these observations are also consistent with the wavelength-dependent features in TPL experiments discussed in the previous section.

Our SNOM operation relies on a high-harmonic filtering procedure (see Methods) that makes a direct evaluation of the FE level impossible[54]. However, by analyzing the results of both SNOM and TPL measurements, we can quantitatively estimate the FE levels in the brightest hot spots by relating their contributions to the overall TPL signal (see Supplementary Note 1). As can be deduced from the experimental near-field spatial distributions (Supplementary Fig. 7), the strongest hot spot within the area of TPL scanning laser spot contributes ~ 20% to the total TPL signal, causing the observed TPL variations (Fig. 1e). By comparing the TPL signals from the bulk and thin gold films and taking into account the size of hot spots measured from the SNOM images, we estimate the average field intensity enhancement of ~ 6000 in hot spots observed with the 5-nm-thin gold film (see Supplementary Note 1). This FE level is favorably compared to that reported in the first experiments on white-light generation in gold dimer antennas[35] and, in general, found consistent with the results reported by other groups for nanostructures containing sub-10-nm-wide gaps (see Supplementary Note 2).

**Discussion**

The main features of the TPL from thin (naturally semi-continuous) near-percolation gold films have been thoroughly investigated using linear reflection spectroscopy along with the TPL and near-field microscopies applied to the films of different thicknesses (ranging from 2 to 11 nm) deposited on glass and silicon substrates. We have mapped the thickness-dependent TPL signals simultaneously with the electrical conductivity measurements, establishing unequivocally that the strongest TPL signals are generated with close to the percolation threshold films. We have revealed the underlying physical mechanisms behind the photo-thermally induced reshaping of nanostructured films that are involved in the polarization and wavelength damage observed in the TPL writing and reading out procedures.

We believe that these easy-to-fabricate and scalable semi-continuous (randomly nanostructured) metallic films near the percolation threshold constitute an important and attractive alternative to more traditional nanostructures, which are admittedly much better defined and controlled but also requiring sophisticated fabrication procedures, used currently for diverse FE applications, including SERS based diagnostics and other kinds of bio-molecule optical sensing. The near-percolation films open up new venues also for direct laser writing in plasmonic nanostructures that can be conducted over wafer-size areas with the capabilities extended by the polarization effect. We estimate the writing beam energy required to write one pixel/bit on the 5-nm-thin gold film to be ∼ 25–50 $\mu$J/bit for glass substrates and ∼ 150 $\mu$J/bit for silicon substrates (Supplementary Fig. 10). Although this level is larger than that obtained with gold nanorods[43], the fabrication of near-percolation films is much more straightforward, while also allowing for multiplexing in depth by sandwiching several thin layers of dielectric and gold. One should not underestimate the importance of exciting possibilities offered by these films for generating broadband and very strong FE effects for quantum plasmonics[4,36-38], especially when taking into account extremely tight confinement of hot spots[55], as revealed by near-field imaging in this work. Finally, white-light generation in the percolation geometry should definitely be investigated further with the perspective of exploiting this phenomenon in nonlinear plasmonics and nanophotonics. Overall, we believe that extremely rich and interesting physics involved in light interaction with near-percolation metal films along with their straightforward and scalable one-step fabrication procedure promises a wide range of fascinating developments and technological applications within diverse areas of modern nanotechnology, from bio-molecule optical sensing to ultra-dense optical data storage.

**Methods**

**Fabrication.** Stripe patterned thin films of gold were fabricated for conductive measurements at different thicknesses by UV- lithography and electron-beam deposition. The borosilicate glass substrates are baked out overnight at 250°C. After baking, the wafers are immediately spin-coated with a 2 μm layer of NZ nLOF 2020 resist, and then exposed with UV-radiation through a shadow mask. After exposure the samples are baked for 2 minutes at 110 °C after which the pattern for the thin films are puddle developed with a 2.38% TMAH water solution. After development, gold is deposited with electron-beam at a vacuum chamber pressure of ∼$10^{-5}$ mbar, and a deposition rate of 2 Å/s, onto the room temperature substrates. After deposition, the samples are transferred to a lift-off bath of Microposit Remover 1165, where they remain for several hours to remove excess photoresist. To protect the thin films, ultrasound is not used during the lift-off process. Films were produced with thicknesses between 2 and 11 nm. The dimensions of the stripes were 1 mm × 250 $\mu$m.

The fabrication was repeated without lithography to produce a set of percolation films on glass with the full range of thicknesses between 2 and 11nm, to avoid any risk of contamination from left-over photoresist for the optical experiments.

Another set of samples on borosilicate glass substrates was prepared to compare bulk gold and thin films. A silicon shadow mask was used to cover one half of the wafer, and a 100 nm bulk gold film was deposited at 10 Å/s using the same deposition system. After removing the mask a 3, 5, 7, or 9 nm thin film was deposited on the wafer with 2 Å/s rate.

The samples on silicon substrate and TEM membranes with thicknesses of 3, 5, 7, and 9 nm were prepared by using the same deposition parameters as the first samples on glass substrates, but no photolithography was used to define the shape of the deposited films.

**SEM and TEM.** For the visualization of gold films, we used a scanning electron microscope Nova NanoSEM from FEI and transmission electron microscope Tecnai T20 G$^2$ from FEI. SEM images were recorded with a through-the-lens detector (TLD), using an acceleration voltage of 3.00 kV at a working distance of 4 mm. TEM images were recorded at an acceleration voltage of 200 kV.

**Linear spectroscopy.** The spectroscopic reflection analysis[30, 31] was performed on a BX51 microscope (Olympus) equipped with a halogen light source, polarizers and a fiber-coupled grating spectrometer QE65000 (Ocean Optics) with a wavelength resolution of 1.6 nm. The reflected and transmitted light was collected using an MPlanFL objective (Olympus) with magnification ×100 (NA =

0.9). The image area analyzed by the spectrometer is limited by a pinhole with a diameter of 150 μm resulting in a circular probing area with a diameter of 1.5 μm. The experimental data in (Supplementary Fig. S1a) represent the reflection ratio $R_{str}/R_{ref}$, where $R_{str}$ is the reflection measured from the films and $R_{ref}$ is the reference from a broadband laser mirror (Edmund Optics, NT64-114) that exhibits an average reflection of 99% between 350 and 1,100 nm of light wavelength.

**Two photon-excited photoluminescence (TPL) microscopy.** We rely on the approach described in Refs. 50–52 and the setup consists of a scanning optical microscope in reflection geometry built on the basis of a commercial microscope and a computer-controlled translation stage. The linearly polarized light beam from a mode-locked pulsed (pulse duration ~200 fs, repetition rate ~80 MHz) Ti-Sapphire laser (wavelength $\lambda$ = 730–860 nm, $\delta\lambda \approx 10$ nm, average power ~300 mW) is used as an illumination source at the FH frequency. After passing an optical isolator (to suppress back-reflection), half-wave plate, polarizer, red colour filter and wavelength selective beam splitter, the laser beam is focused on the sample surface at normal incidence with a Mitutoyo infinity-corrected long working distance objective (×100, NA = 0.70). The half-wave plate and polarizer allow accurate adjustment of the incident power. TPL radiation generated in reflection and the reflected FH beam are collected simultaneously with the same objective, separated by the wavelength selective beam splitter, directed through the appropriate filters and detected with two photomultiplier tubes (PMTs). The tube for TPL photons (within the transmission band of 350–550 nm) is connected with a photon counter giving typically only ~20 dark counts per second (cps). The FH and TPL spatial resolution at full-width-half-maximum is ~0.75 μm and ~0.35 μm, respectively, which means no individual clusters will be resolved in the TPL images. In this work, we used the following scan parameters: the integration time (at one point) of 50 ms, scanning speed (between the measurement points) of 20 μm/s, and scanning step sizes of

~350 and 700 nm. We adjusted the incident power $P$ within the ranges of 0.15–0.5 mW, for films on glass substrates, and 0.5–1.2 mW, for films on silicon substrates, in order to obtain significant TPL signals and record the TPL signal dependence on the incident powers. For writing we used incident power $P$ within the ranges of 0.6-4 mW and 2.5-5 mW for films on glass and silicon substrates, respectively. For the reference bulk gold sample, we confirmed that the TPL signals obtained depend quadratically on the incident power. During these measurements, we kept for simplicity the excitation wavelength fixed at 740 nm, since there are no specific resonances (Supplementary Fig. 1a) at this wavelength, and 740 nm is more visible and convenient to focus.

**Electrical measurements.** For the conductive measurements, we used the two-probe method and results obtained by a Keithley 2400 SourceMeter. In the experiment, we used the stripes on silica substrate (see fabrication) with an electrical contact made at each end. Electrical contacts were established using EPO-TEK H20E conductive epoxy.

**Photoluminescence spectroscopy.** To record spectra of the observed photoluminescence, we used the same setup as for TPL measurements but instead of the PMT for TPL photons we used the spectrometer QE65000 (Ocean Optics) and with only a filter to cut off the laser line. Since long exposure time or high power could damage the sample during recording of photoluminescence spectra, we continuously scanned the sample with the following parameters. Integration time (at one point) of 50 ms, scanning speed (between the measurement points) of 20 μm/s, scanning step size of ~350 nm, and incident power P ~ 0.4 mW for gold film on silica substrate and ~ 2 mW on silicon. The recording time for the spectrum is 60 s.

**Near-field microscopy.** The near-field investigations were performed using a scattering-type SNOM based on an atomic force microscope (AFM), in which the near-field is scattered by an uncoated silicon probe, operating in a tapping mode at a frequency $\Omega \approx 250$ kHz. The sample was illuminated normally from below (transmission mode[54]) with a linearly polarized tuneable (1425-1625 nm) telecom laser. The scattered signal was detected and demodulated at the fourth harmonic $4\Omega$ to filter the near-field contribution from the background. Additionally, the interferometric pseudoheterodyne detection[56] was employed, which allows imaging of both the amplitude and the phase of the near-field.

**Acknowledgments:** The authors gratefully acknowledge financial support from the European Research Council, Grant 341054 (PLAQNAP), CONACyT Basic Scientific Research Grant 250719, and the Danish Council for Independent Research–Natural Sciences (Project 1323-00087). Center for Nano Optics was financially supported from the University of Southern Denmark (SDU 2020 funding), while Center for Nanostructured Graphene (CNG) was funded by the Danish National Research Foundation (CoE Project DNRF103).

**Author contributions:** S. I. B. and J. B. conceived the experiment. C. F. and N. S. designed and fabricated the samples and performed the electron microscopy (SEM and TEM). S. M. N. and V. C. performed the two-photon luminescence experiments. S. M. N. conducted electrical measurements and optical spectroscopy. V. A. Z. performed and analyzed near-field microscopy measurements. S. M. N. and J. B. drafted the manuscript. All authors discussed the results and commented on the manuscript. S. I. B. and N. A. M. supervised the project.

**Conflict of interests:** The authors declare no competing financial interests.

**Correspondence:** Correspondence and requests for materials should be addressed to S. I. B. (email: seib@iti.sdu.dk).

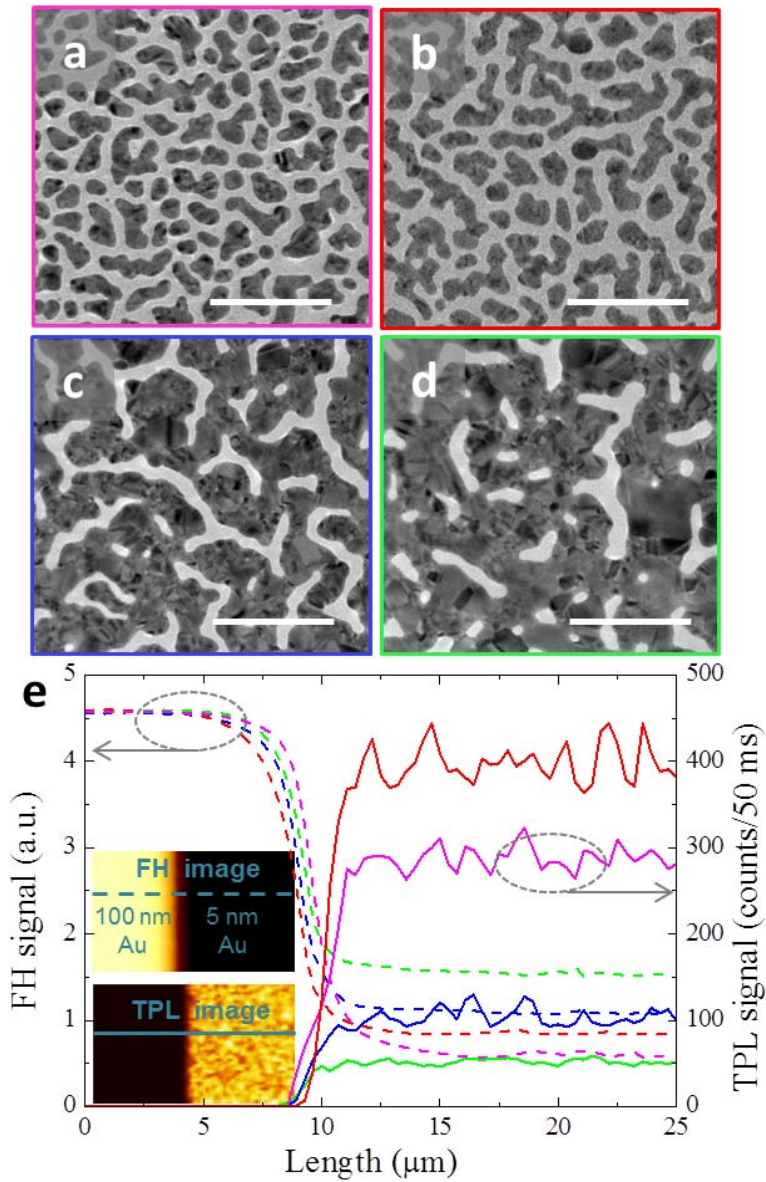

**Figure 1: Gold near-percolation films.** TEM images of thin gold films with nominal thicknesses of (**a**) 3, (**b**) 5, (**c**) 7, and (**d**) 9 nm. Dark grey corresponds to the gold. The scale bars are 100 nm. (**e**) Cross sections of FH (dashed line) and TPL signal (solid line) across the border of bulk gold and thin gold film with nominal thickness of 3 nm (magenta), 5 nm (red), 7 nm (blue), and 9 nm (green). The average incident power was ~0.4 mW. Insets: typical FH (upper) and TPL (lower) images, with lines indicating the orientation of cross sections.

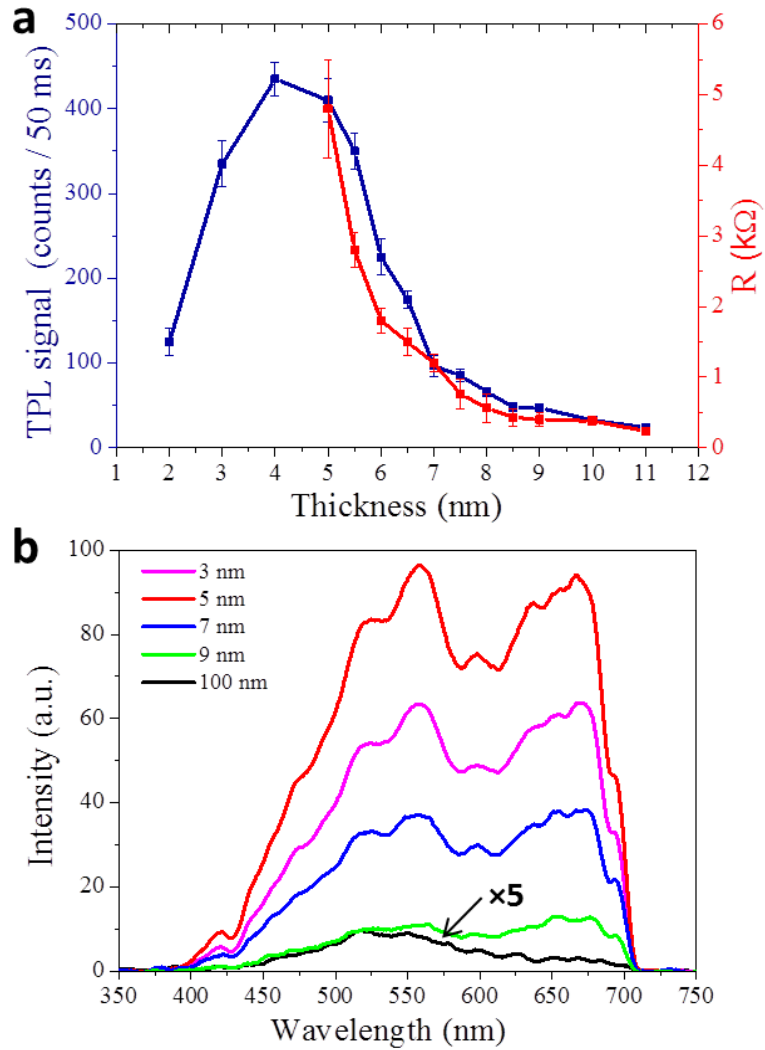

**Figure 2: TPL correlation with percolation and luminescence spectra.** (**a**) The TPL signal (blue) and resistance (red) measured for gold films with thicknesses from 2 to 11 nm at the excitation wavelength of 740 nm. (**b**) Luminescence spectra obtained for the gold films of thicknesses 3 nm (magenta), 5 nm (red), 7 nm (blue), 9 nm (green), and 100 nm (black) at the excitation wavelength of 780 nm. The luminescence intensity from bulk gold is multiplied by 5 for visibility. Arbitrary unit in (**b**) corresponds to 200 counts/s. The average incident power was ~ 0.4 mW for thin gold films and ~ 10 mW for bulk gold.

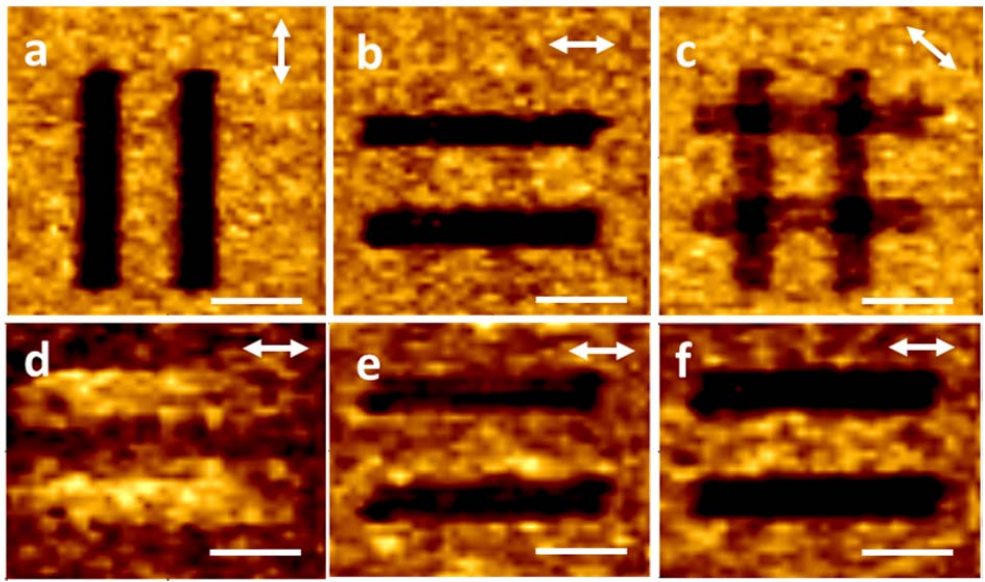

**Figure 3: TPL polarization- and wavelength-dependent read-out.** TPL images obtained from the 5-nm-thin gold film on a silicon substrate, which demonstrate (**a**)-(**c**) polarization and (**d**)-(**e**) wavelength dependences. Two pairs of lines in (**a**)-(**c**) are written with the FH polarization along the corresponding lines. Both writing and reading were carried out at the wavelength of 740 nm. A pair of lines in (**d**)-(**f**) is written at the wavelength of 780 nm with the FH polarization along the correspondent lines, while the reading was done at the wavelength of (**d**) 740, (**e**) 780, and (**f**) 820 nm. Double arrows indicate the FH polarization during the read-out. The FH beam powers of ~ 3 and ~ 1 mW are used for writing and reading, respectively. The scale bars are 5 μm.

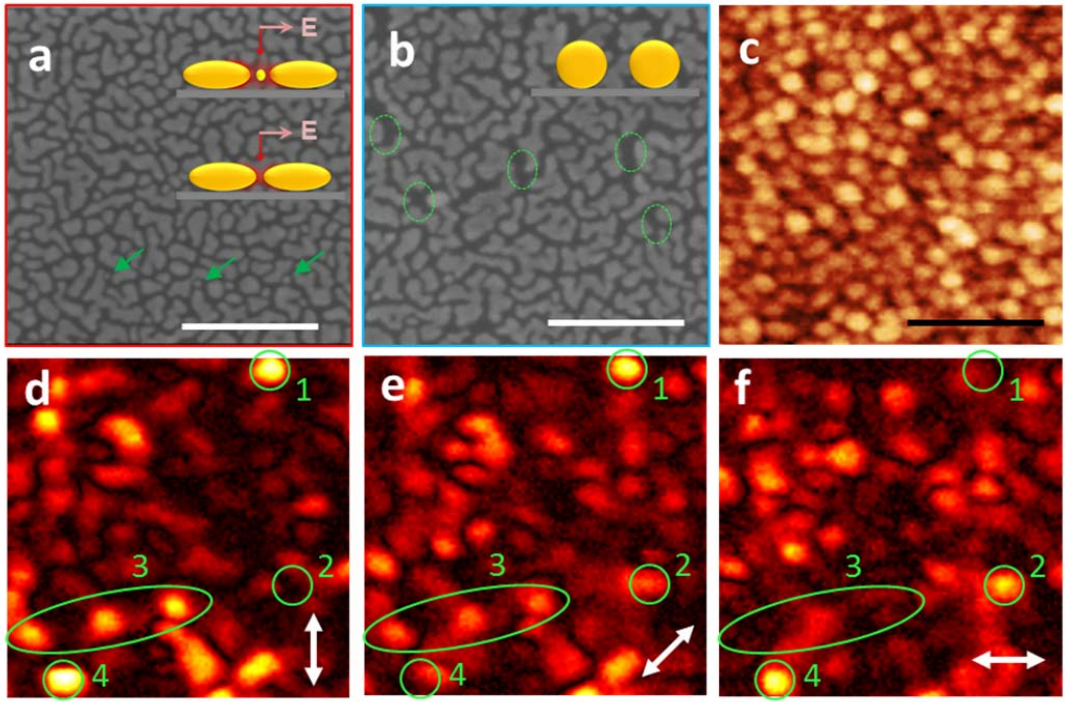

**Figure 4: Spatial anisotropy.** SEM images of (**a**) the pristine 5-nm-thin gold film on a silicon substrate and (**b**) the same film after the laser-induced damage. Insets schematically depict the process, where hot spots created inside a small gap with or without a small gold particle, heat up under illumination, resulting in melting and reshaping, eventually increasing the gap. Green arrows in (**a**) show some of the small particles presented in the undamaged sample, and green ellipses in (**b**) represent large gaps created after laser illumination. (**c**)-(**f**) Pseudocolour SNOM images of (**c**) topography and (**d**)-(**f**) optical near-field amplitude of the pristine 5-nm-thin gold film on a glass substrate. Double arrows represent the illumination polarization of the telecom laser ($\lambda = 1500$ nm). Green circles in (**d**)-(**f**) encircles the same hot spots. The scale bars in (**a**)-(**c**) are 200 nm.

# White light generation and anisotropic damage in gold films near percolation threshold: Supplementary information

Sergey M. Novikov[1], Christian Frydendahl[2,3], Jonas Beermann[1], Vladimir A. Zenin[1], Nicolas Stenger[2,3], Victor Coello[4], N. Asger Mortensen[2,3] & Sergey I. Bozhevolnyi[1]

[1]*Centre for Nano Optics, University of Southern Denmark, Campusvej 55, DK-5230 Odense M, Denmark*

[2]*Department of Photonics Engineering, Technical University of Denmark, DK-2800 Kongens Lyngby, Denmark*

[3]*Center for Nanostructured Graphene, Technical University of Denmark, DK-2800 Kongens Lyngby, Denmark*

[4]*CICESE Monterrey, Alianza Centro No.504 PIIT Apodaca, N. L. C. P. 66600, Mexico*

\* To whom correspondence should be addressed
E-mail: seib@iti.sdu.dk

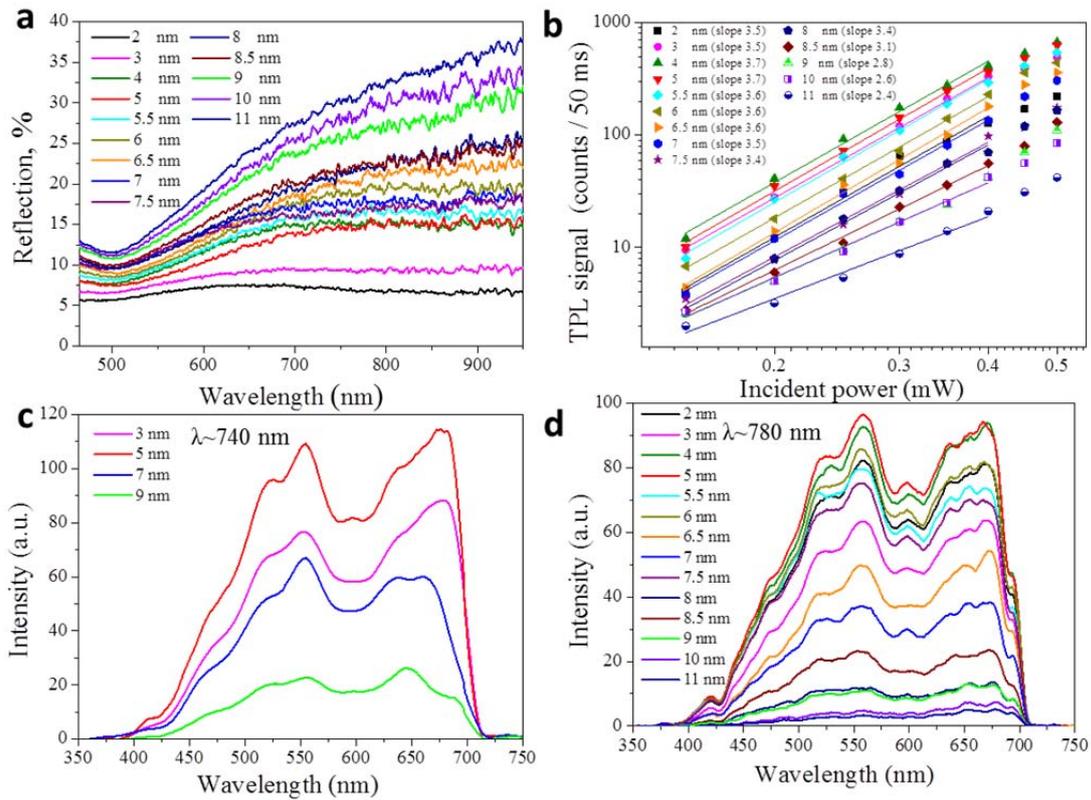

**Supplementary Figure 1: Characterization of thin gold films on glass substrate.** (**a**) Reflection spectra obtained for the thin gold films and normalized as explained in Methods. (**b**) Dependence of TPL signal on the incident power. The data (points) were fitted with lines, whose slope indicates the order of TPL dependence on the power. The fitting was done until the power of 0.4 mW, because higher power might cause damage. Luminescence spectra obtained at the excitation wavelength of (**c**) 740 and (**d**) 780 nm. The arbitrary unit used in (**c**) and (**d**) corresponds to 10 counts/50 ms or 200 counts/s, respectively.

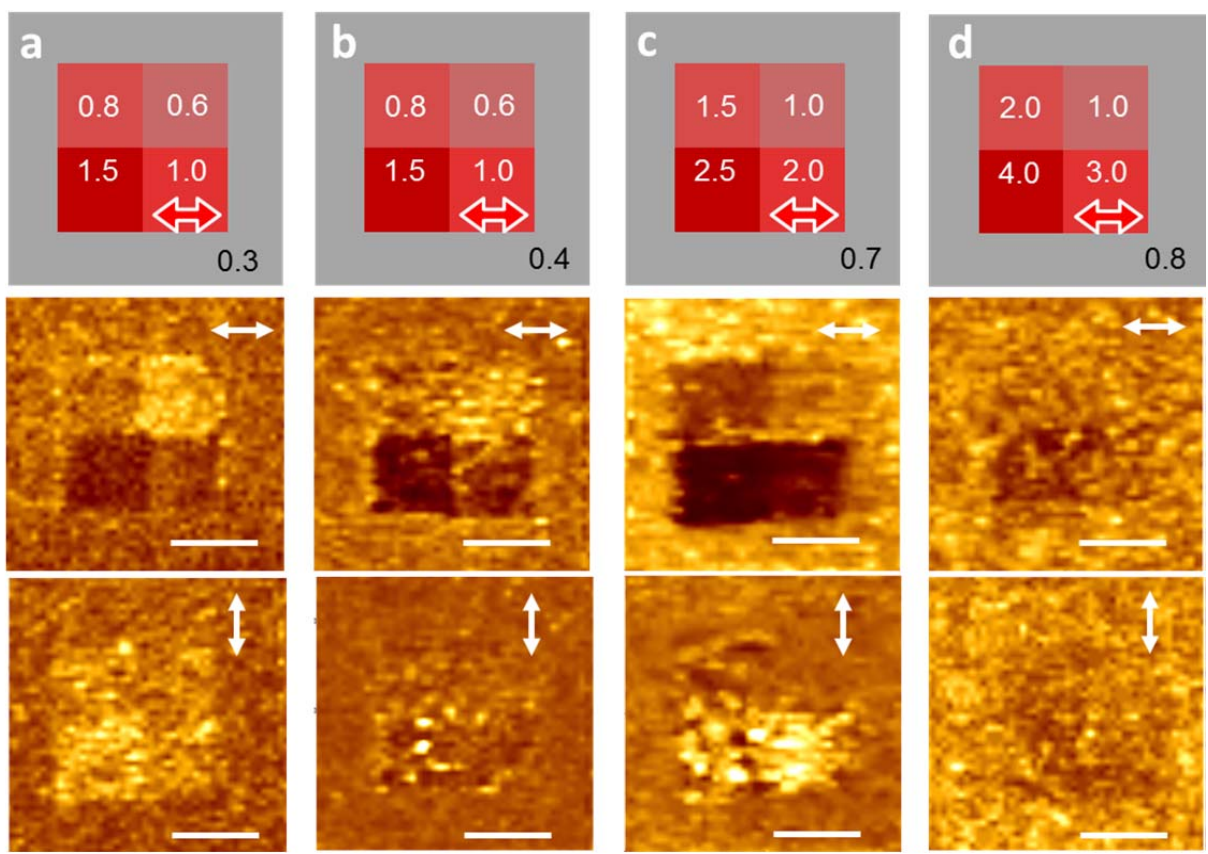

**Supplementary Figure 2: Polarization dependent TPL imaging.** TPL images obtained for thin gold films on glass substrates with film thickness of (**a**) 3, (**b**) 5, (**c**) 7 and (**d**) 9 nm. Top: a schematic view of written squares (red) with numbers indicating used the writing laser power in mW. Numbers in the low right corner of grey square indicate the laser power in mW used for the read-out. Arrows indicate the electric field polarization direction of the scanning laser used for writing (top) and reading (central and bottom rows). The scale bars are 5 μm.

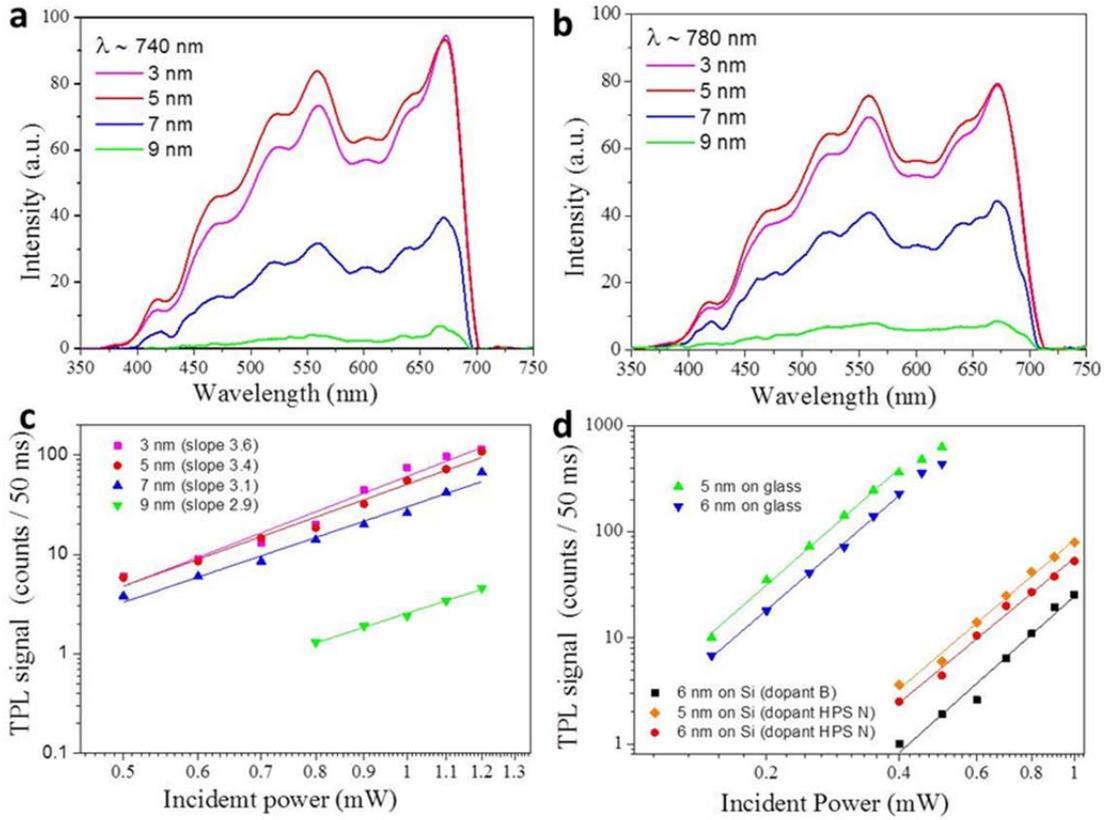

**Supplementary Figure 3: Substrate influence on the TPL.** Luminescence spectra obtained for the excitation wavelength of (**a**) 740 and (**b**) 780 nm for thin gold films with thicknesses of 3 (magenta), 5 (red), 7 (blue) and 9 nm (green) on silicon substrates. (**c**) Dependences of TPL signals on the incident power for gold films on silicon substrates. The data (points) were fitted with lines, whose slope indicates the order of TPL dependence on the power. (**d**) TPL power dependences for the 5- and 6-nm-thin gold films deposited on glass and silicon (with different dopants) substrates. The arbitrary units used in (**a**) and (**b**) correspond to 10 counts/50 ms or 200 counts/s.

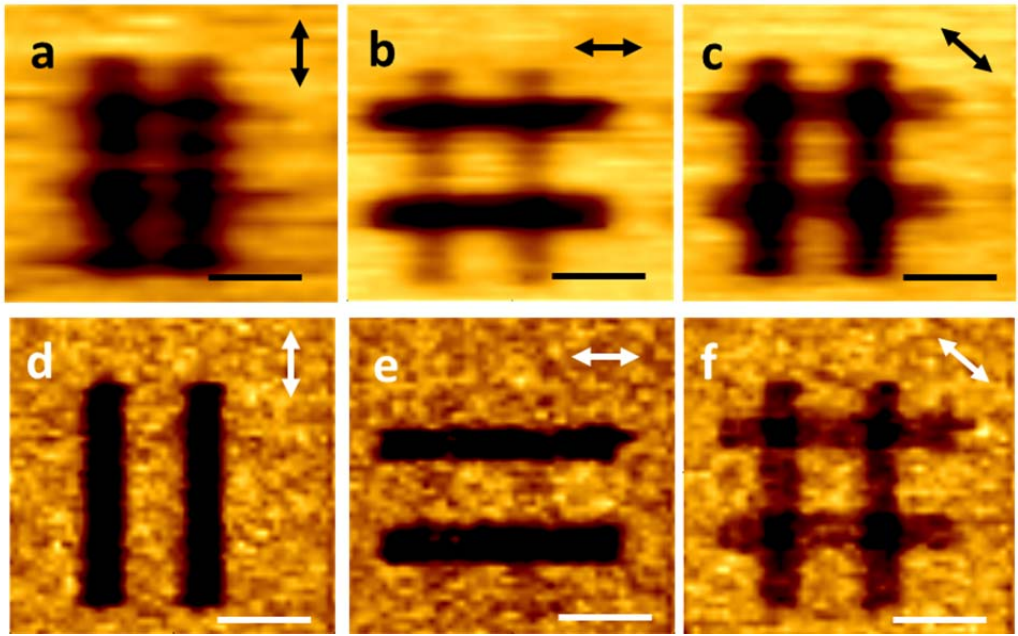

**Supplementary Figure 4: Comparing FH and TPL images during the read-out.** (**a**)-(**c**) FH and (**d**)-(**f**) TPL images obtained from the 5-nm-thin gold film on a silicon substrate upon reading with the polarization direction indicated with a double arrow. Both writing and reading are carried out at the wavelength of 740 nm. The scale bars are 5 μm.

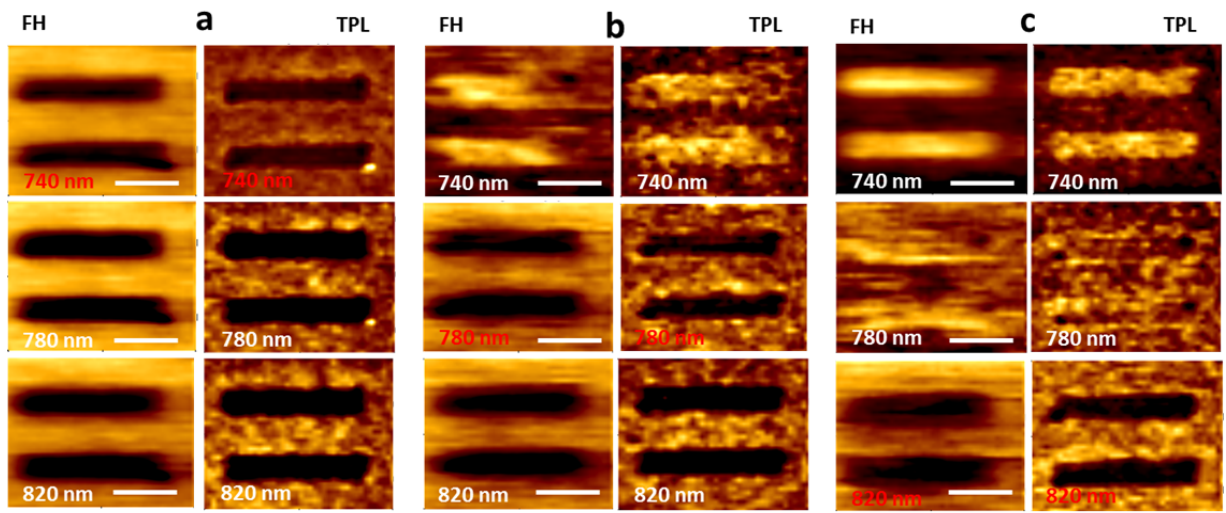

**Supplementary Figure 5: Wavelength dependent TPL imaging.** The fundamental harmonic (left) and TPL (right) images obtained from the 5-nm-thin gold film on a silicon substrate. The writing was carried out at the wavelength of (**a**) 740, (**b**) 780, and (**c**) 820 nm. The FH wavelength used for reading was 740 nm (top row), 780 nm (middle row), and 820 nm (bottom row). Both writing and reading were conducted with the FH polarization direction along the written lines. The scale bars are 5 μm.

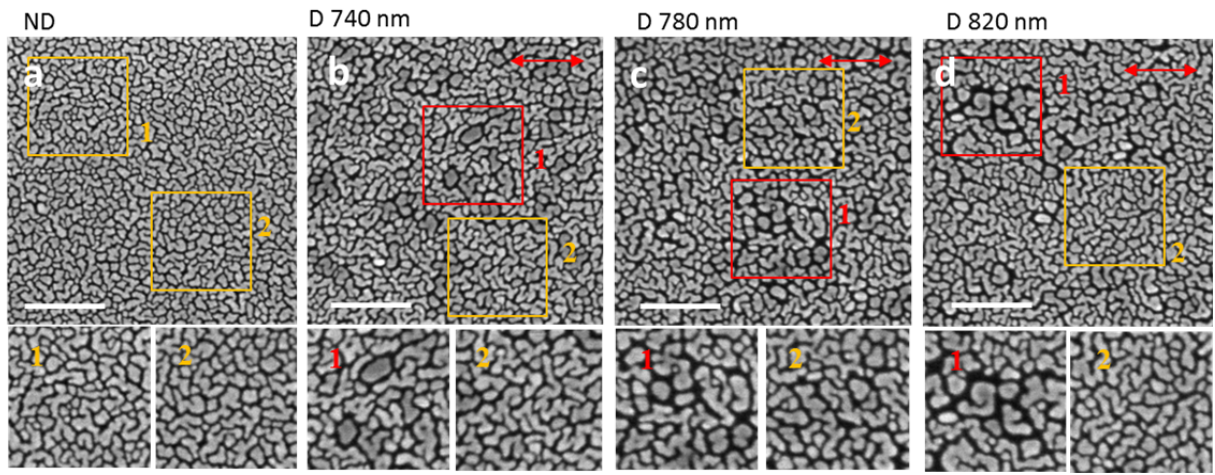

**Supplementary Figure 6: SEM characterization.** SEM images obtained with the 5-nm-thin gold film on a silicon substrate (**a**) before and after damage inflicted at (**b**) 740, (**c**) 780, and (**d**) 820 nm. Orange frames indicate areas without noticeable damage while red frames show areas with melted clusters (bottom panels show enlarged frames). The scale bars are 250 nm.

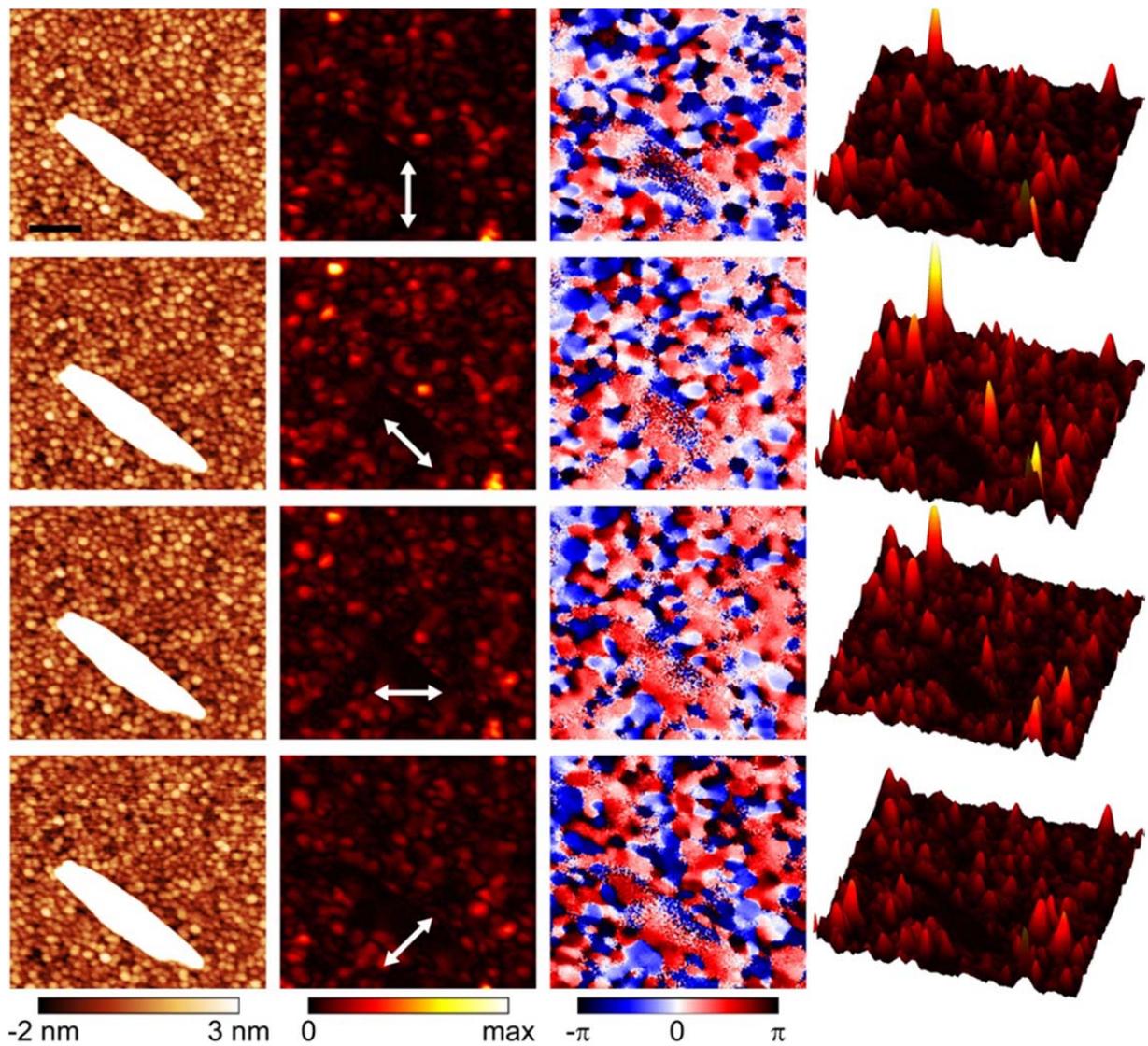

**Supplementary Figure 7: Polarization dependent SNOM images.** Pseudocolour SNOM images of topography (left), optical near-field amplitude (middle) and phase (right) of the pristine 5-nm-thin gold film on a glass substrate. Right most column shows amplitude maps in 3D. Double arrows represent the illumination polarization of the telecom laser ($\lambda$ = 1500 nm). White mark (visible in topography) was used for alignment. The scale bar is 200 nm.

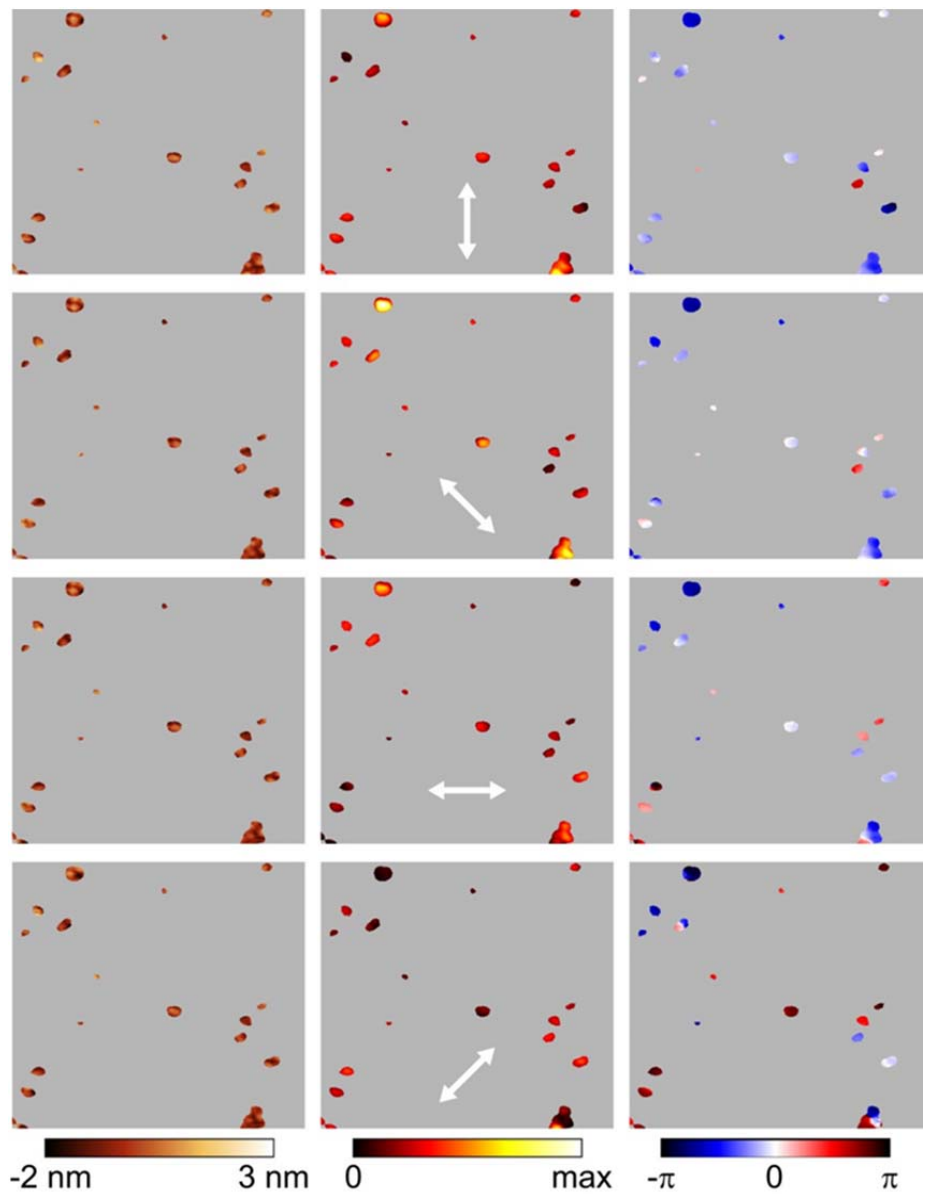

**Supplementary Figure 8: Identification of hot spots.** The same as in Supplementary Fig. 7 SNOM images of topography (left), optical near-field amplitude (middle) and phase (right) of the pristine 5-nm-thin gold film on glass substrate, masked by the same mask. The mask was made from the average near-field amplitude (for all polarizations) to highlight hot spots. Double arrows represent the illumination polarization of the telecom laser ($\lambda = 1500$ nm).

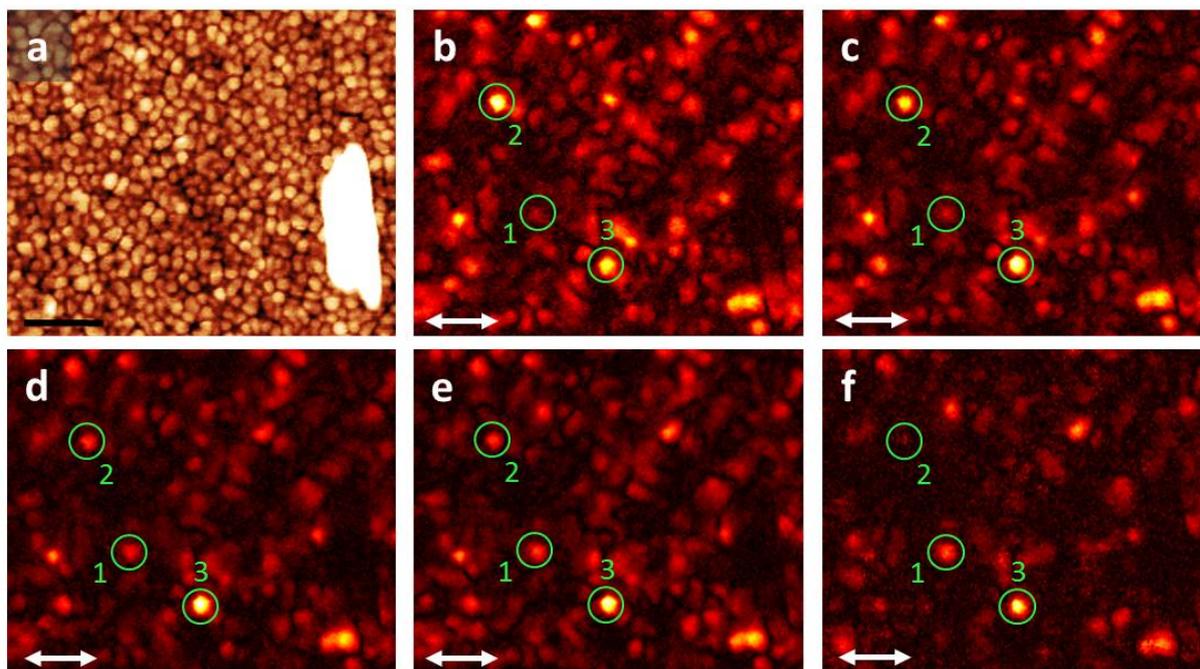

**Supplementary Figure 9: Wavelength dependent SNOM images.** Pseudocolour SNOM images of (**a**) topography and (**b**)-(**f**) optical near-field amplitude of undamaged 5-nm-thin gold film on glass substrate at the illumination wavelength of (b) 1425, (**c**) 1475, (**d**) 1525, (**e**) 1575 and (**f**) 1625 nm, respectively. Double arrows represent the illumination polarization, kept constant for all wavelengths. White mark (visible in topography) was used for alignment. Colour maps are the same as in Supplementary Fig. 7. Each near-field amplitude image was normalized to the maximum at the given illumination wavelength. The scale bar is 200 nm. Green circles in (**b**)-(**f**) encircle the same hot spots.

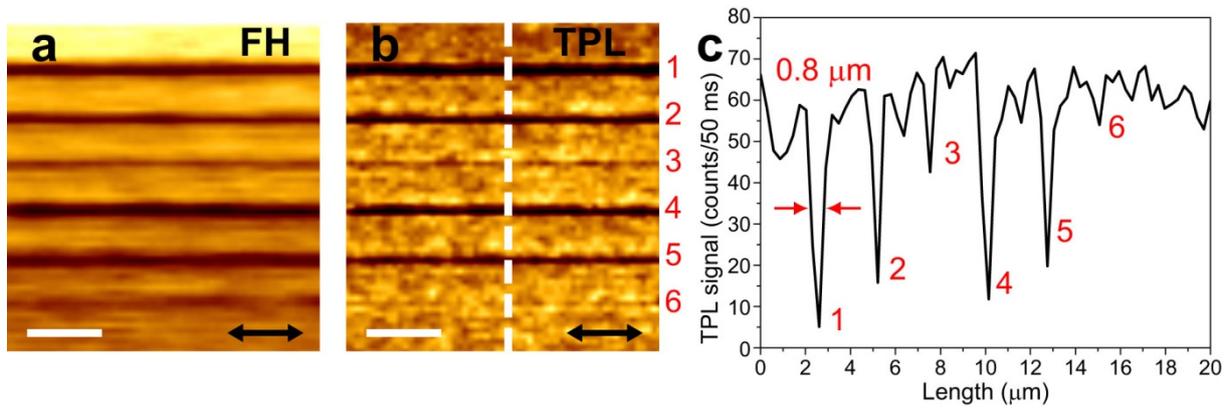

**Supplementary Figure 10: Estimation of the energy required for writing.** (**a**) FH and (**b**) TPL images of the 5-nm-thin gold film on a silicon substrate, obtained with different FH powers during the writing procedure (with the exposure time of 50 ms and the scanning speed of 20 μm/s) and different scanning steps. For the scanning step of 360 nm, the FH writing powers are $P_1$ ~5 mW, $P_2$ ~4 mW, $P_3$ ~3 mW, whereas for the scanning step of 800 nm, $P_4$ ~5 mW, $P_5$ ~4 mW, $P_6$ ~3 mW. (**c**) Cross sections taken from the place marked by the dashed line in (**b**). The scale bars are 5 μm.

**Supplementary discussion**

**1. Estimation of the field enhancement**

Let us consider the origin of two-photon luminescence (TPL) signals measured with both optically thick and very thin, island-like, gold films. We assume that the TPL signal is produced inside gold and leaves gold nanostructures without much absorption, since gold becomes increasingly transparent for wavelengths shorter than 600 nm.[1] TPL mechanism implies that the TPL power depends quadratically upon the incident light power or, alternatively expressed, upon the fourth power of the electric field magnitude, $|E_g|$, inside the gold:

$$P_{TPL} = \xi \int_V |E_g|^4 dV , \qquad (1)$$

where $\xi$ is a proportionality constant depending on gold properties and the illumination wavelength, and the integration is performed only inside the gold volume.

Considering the TPL signal from optically thick gold films (or bulk gold), we first express the electric field inside the gold using Fresnel relations and assuming normal incidence and relatively weak focusing:

$$E_g(x,y,z) = \frac{2}{n_g + 1} E(x,y) \exp\left[i \frac{2\pi}{\lambda} n_g z\right] , \qquad (2)$$

where $E(x,y)$ is the incident electric field amplitude at the gold surface ($z$ is the depth coordinate) and $n_g$ is a complex-valued refractive index of gold: $n_g = 0.1686 + 4.5824i$ at the light wavelength $\lambda = 750$ nm.[1] The TPL power from the bulk gold can then be found using Eqs. (1) and (2) as follows:

$$P_{TPL}^b = \frac{16}{(|n_g + 1|)^4} \xi \int_V |E(x,y)|^4 dxdy \left|\exp\left[i\frac{2\pi}{\lambda}n_g z\right]\right|^4 dz =$$

$$= \left\|\begin{matrix} n_{gi} = \text{Im}(n_g) \gg \text{Re}(n_g) \\ \delta_{TPL} = \lambda/8\pi n_{gi} \end{matrix}\right\| \cong \frac{16\xi}{n_{gi}^4} \delta_{TPL} \int_S |E(x,y)|^4 dxdy \qquad (3)$$

with $\delta_{TPL}$ characterizing the TPL skin depth, which amounts to ~ 6.5 nm at $\lambda = 750$ nm. In evaluating the depth integral, we have implicitly assumed the film thickness to significantly exceed $\delta_{TPL}$. Introducing the average magnitude $E_{in}$ of the incident electric field and the incident beam spot area $S$, allows us to finally obtain:

$$P_{TPL}^b \cong \frac{16\xi}{n_{gi}^4} \delta_{TPL} E_{in}^4 S \quad . \tag{4}$$

Considering the TPL signal from thin, island-like, gold films, we disregard depth variations (i.e., along the $z$ coordinate) of the incident electric field inside gold films to simplify the analysis (also the films are thinner than 10 nm). Our analysis of near-field images with the size similar to that of the incident laser beam in TPL experiments (see Fig. 4 and Supplementary Fig. 8) indicates that the integration of the fourth power of the near-field amplitude over the whole image area and only over the brightest in the image spot (hot spot) produces the results different by ~ 5 times:

$$p \int_S |E_{NF}|^4 \, dxdy = \|p \cong 0.2\| = \int_{hot\ spot} |E_{NF}|^4 \, dxdy \cong E_{max}^4 S_{hs} \quad , \tag{5}$$

where $E_{max}$ is the maximum magnitude of the electric field (in the strongest hot spot) and $S_{hs} \cong (30\ \text{nm})^2$ is a typical size of the hot spot in near-field images. During the TPL imaging, the illuminating laser beam would occasionally miss the brightest hot spot available resulting in the variations of the detected TPL signal on the level of 10% (see Fig. 1**e**), which is consistent with Eq. (5) deduced from the near-field images. We further note that the strongest electric field is expected to occur inside narrow gaps between gold islands (as is discussed in detail in the main text and corroborated with images in Fig. 4). For gaps being much smaller than the free-space wavelength, the electric field inside the gap is oriented predominantly perpendicular to the gap boundaries and enhanced due to the boundary conditions.[2] The latter allows us to relate the strongest field in the gap, $E_{max}$, to the corresponding field inside the gold, $E_{g\,max}$, that generates the TPL:

$$E_{max} = \left| n_g^2 E_{g\,max} \right| \Rightarrow E_{g\,max} \cong \frac{E_{max}}{n_{gi}^2} \quad . \tag{6}$$

Assuming additionally that the relation expressed by Eq. (5) can also be applied to the field inside the gold, we can relate the TPL power measured with a film of thickness $t$ to the maximum gap electric field:

$$P_{\text{TPL}}^f = \xi \int_V |E_g|^4 dV = \xi t \int_S |E_g|^4 dxdy = \xi t \frac{1}{p} E_{g\max}^4 S_{\text{hs}} = \frac{\xi t}{p} \left( \frac{E_{\max}}{n_{gi}^2} \right)^4 S_{\text{hs}} . \quad (7)$$

The obtained formulae, Eq. (4) and Eq. (7), allow us to estimate the electric field intensity enhancement, $\alpha$, in the strongest hot spots of a given sample (with respect to the incident field intensity) by relating the TPL power measured with the sample and with an optically thick gold film (the reference sample) for the same illuminating beam:

$$\alpha = \frac{E_{\max}^2}{E_{in}^2} \cong 4 n_{gi}^2 \sqrt{\frac{P_{\text{TPL}}^f}{P_{\text{TPL}}^b}} \sqrt{\frac{p \delta_{\text{TPL}} S}{t S_{hs}}} . \quad (8)$$

The first factor in the above formula comes from the difference in the relations between the fields inside and outside gold [cf. Eq. (2) and Eq. (6)]. The second factor takes into account the difference in the TPL signals measured, while the third one compares the effective volumes of gold from which TPL signals originate. Often the TPL measurements are conducted with incident laser beams of different powers that are adjusted individually so as to have sufficiently strong TPL signals without damaging illuminated regions. Introducing the powers of incident beams for the considered cases as $P_{\text{FH}}^b$ and $P_{\text{FH}}^f$, respectively ("FH" stands for "fundamental harmonic" for historical reasons[3]), obtains the final expression:

$$\alpha = \frac{E_{\max}^2}{E_{in}^2} \cong 4 n_{gi}^2 \sqrt{\frac{P_{\text{TPL}}^f}{P_{\text{TPL}}^b} \frac{P_{\text{FH}}^b}{P_{\text{FH}}^f}} \sqrt{\frac{p \delta_{\text{TPL}} S}{t S_{hs}}} . \quad (9)$$

Considering the TPL measurements conducted with a 5-nm-thin gold film on glass, the contribution from the first factor is ~ 85, while the ratio of the TPL powers normalized to the same incident power is ~ 80 ($P_{\text{TPL}}^b \cong 7$ counts/50 ms $\cong$ 140 counts/s at $P_{\text{FH}}^b = 10$ mW, $P_{\text{TPL}}^f \cong 10$ counts/50 ms $\cong$ 200 counts/s at $P_{\text{FH}}^f = 0.15$ mW). For the same 5-nm-thin gold film on glass, taking $p = 0.2$, $\delta_{\text{TPL}} \cong 6.5$ nm, $S = (500 \text{ nm})^2$, $t = 5$ nm, $S_{hs} \cong (30 \text{ nm})^2$, the third factor is ~ 8.5. All in all, the intensity enhancement factor is estimated as $\alpha \sim 6 \cdot 10^4$.

Considering various assumptions involved in our estimation of the intensity enhancement [Eq. (9)], probably the most important one is related to the usage of the same proportionality constant, $\xi$, in both cases [cf. Eq. (1) and Eq. (7)]. This assumption disregards the influence of configuration geometry on the TPL process, which is essentially the process of photon emission with its rate being influenced by the local density of states (that can also be expressed by the Purcell factor). This implies that the TPL signal can strongly be enhanced by resonant characteristics of a gold nanostructure at the emission wavelength,[4,5] whereas our consideration takes into account the field enhancement only at the wavelength of illumination [Eq. (5)]. It is clear that, since we attempt on isolating the intensity enhancement in the *brightest* hot spots, we should incorporate also this mechanism of TPL enhancement. While it is very challenging to theoretically describe this effect for irregular island-like films, one can make use of the available experimental results on the TPL enhancement caused by particle resonances at the TPL wavelength and estimate this TPL power enhancement as < 100.[4,5] Decreasing correspondingly the TPL power used in Eq. (9), results in the *final* (conservative) estimate of the field intensity enhancement: $\alpha \sim 6 \cdot 10^3$.

## 2. Comparison with other configurations

Let us compare the obtained estimate of the intensity enhancement, $\alpha \sim 6 \cdot 10^3$, with those reported in literature. White light generation and non-quadratic TPL power dependences were also observed for dimer gold antennas.[6] For this configuration, the authors calculated the maximum field intensity enhancement in the plane located 10 nm above gold antennas as ~ 200 for the gap of 30 nm. Without going into details of their estimation procedure, we note that the gap width influences the size of a hot spot, and, in our case, is expected to be *considerably* smaller, at least for the films that are near the percolation threshold. In a very recent work devoted to radiation nanofocusing by using tapered metal stripes terminated with impedance-matched resonant antennas,[7] the estimated intensity enhancement of $\sim 12 \cdot 10^3$ was found inside a 10-nm-wide gap (with a square cross-section of 30×30 nm$^2$). Even though this number is close to our estimate, we should note that these configurations are very different because, in the referred work,[7] the average electric field inside the gap was compared to the average electric field of the propagating (along a metal stripe) plasmonic mode. A better number for comparison can be found in accompanying Supplementary Figure S8, where the characteristics of a resonant dimer antenna with the same gap are displayed, indicating the intensity enhancement of $\sim 9\cdot 10^3$.[7] Even larger intensity enhancements, reaching $\sim 11\cdot 10^3$, were reported when studying silver dimers consisting of 36-nm-diameter spheres with a 2-nm-wide gap, while the intensity enhancement of $\sim 5\cdot 10^4$ was estimated for the same gap of 2 nm between two 12-nm-thin silver triangles having a side length of 60 nm in a bow-tie configuration.[8] Note that, in this work, the estimations were conducted using the classical electrodynamics with a local response. Theoretical calculations based on a more accurate approach involving a nonlocal response predict the enhancement of $\sim 10^4$ inside a 1-nm-wide gap between two silver cylinders with radii of 15 nm.[9] Therefore, our estimate of the intensity enhancement, $\alpha \sim 6\cdot 10^3$, seems to be a reasonable value, also because, for very thin, island-like gold films near the percolation threshold all different gap widths are expected.

**Supplementary References**